# Comment on "Linear and passive silicon optical isolator" in Scientific Reports 2, 674

Alexander Petrov,[1,*] Dirk Jalas,[1] Manfred Eich,[1] Wolfgang Freude,[2,3] Shanhui Fan,[4] Zongfu Yu,[4] Roel Baets,[5,6] Miloš Popović,[7] Andrea Melloni,[8] John D. Joannopoulos,[9] Mathias Vanwolleghem,[10] Christopher R. Doerr,[11] Hagen Renner,[12]

[1] Institute of Optical and Electronic Materials, Hamburg University of Technology, D-21073 Hamburg, Germany
[2] Institute of Photonics and Quantum Electronics (IPQ), Karlsruhe Institute of Technology (KIT), 76131 Karlsruhe, Germany
[3] Institute of Microstructure Technology (IMT), Karlsruhe Institute of Technology (KIT), 76131 Karlsruhe, Germany
[4] Ginzton Laboratory, Department of Electrical Engineering, Stanford University, Stanford, CA 94305, USA
[5] Photonics Research Group, INTEC-department, Ghent University-IMEC, B-9000 Gent, Belgium
[6] Center for Nano- and Biophotonics, Ghent University, B-9000 Gent, Belgium
[7] Department of Electrical, Computer and Energy Engineering, University of Colorado, Boulder, CO 80302, USA
[8] Dipartimento di Elettronica e Informazione, Politecnico di Milano, 20133 Milano, Italy
[9] Department of Physics, Massachusetts Institute of Technology, Cambridge, MA 02139, USA
[10] Institut d'Electronique de Micro-electronique et de Nanotechnologie, CNRS, Université Lille 1, 59650 Villeneuve d'Ascq, France
[11] Acacia Communications Inc., Three Clock Tower Place, Suite 210, Maynard, MA 01754, USA
[12] Institute of Optical Communication Technology, Hamburg University of Technology, D-21073 Hamburg, Germany
*Corresponding author: a.petrov@tuhh.de

Wang et al. [1] demonstrated different power transmission coefficients for forward and backward propagation in simulation and experiment. From such a demonstration, the central claim of their paper is that "the spatial inversion symmetry breaking diode can construct an optical isolator in no conflict with any reciprocal principle". Their claim contradicts the Lorentz reciprocity theorem, from which it is well known that one cannot construct an isolator this way.

In their article Wang et al. [1] claim to achieve optical isolation in purely linear and passive silicon photonics structures in no conflict with the Lorentz reciprocity theorem. We believe that the authors were unaware of the fact that higher order modes of the considered waveguide sections contributed significantly to the outcome of their simulations and measurements. In addition, simulated and experimental results cannot be compared if the excited modes are not defined. Specifying only the total power flowing over the cross section of a waveguide does not unambiguously determine the excitation and transmission conditions in a multimode waveguide. It may well occur that for certain excitation conditions the total power flow in the waveguide cross section is different for forward and backward directions. But this fact does not allow building an optical isolator, because for the very same structure there are also excitation conditions for which forward and backward transmitted powers are equal. A simple example of such a device is the coupling between a single mode and a multimode fibre. There will be almost 100% power transmission from a single mode fibre into a multimode fibre but, vice versa, during backward propagation the modes of the multimode fibre -if not excited with exactly the same phase and amplitude relations as resulted from the previous single mode fibre excitation- will be strongly scattered and reflected by the single mode fibre. Such a simple arrangement can by no means be called an optical isolator and cannot implement the functionality of an optical isolator. The reason is that a proper amplitude and phase combination of the modes in the multimode fibre will lead to almost 100% backward power transmission. It should be also said that the complex coupling coefficients between the mode of a single mode fibre in one of two orthogonal polarisations (representing one port each) to a particular mode of the multimode fibre in a specific polarisation (representing another port) are the same for forward and backward propagation. Thus even though the overall power transmission from a single mode fibre to a multimode fibre is almost 100%, the transmission into any particular mode of the multimode fibre is usually very low.

The authors have cited our Comment in Science Journal [2], which corrected a flaw in a paper published in Science. Unfortunately, they are making exactly the same kind of mistake as was made in the Science paper we commented. Their explanation that the obtained results don't contradict the reciprocity principle is incorrect. In their derivation the authors label as "port A" and "port B" multimode waveguides, whereas, for reasons stated above, each mode for each polarization should have been labeled as a separate port. But even if their waveguides were single moded, the derivation provided is still incorrect. The authors apply time-reversal symmetry and confuse it with reciprocity. It should be mentioned here, that reciprocity does not require time-reversal symmetry. An example would be a system with absorption loss. The reversed transmitted signals on their way back through the structure will be absorbed a second time and will never result in the initial excitation. Thus, time reversal symmetry is not fulfilled in this case. At the same time, the system with loss is still reciprocal and has equal transmission coefficients for forward and backward propagation. In the article the authors base their explanation on time reversal symmetry. By excluding the scattered energy they come to the conclusion that the reversed signal will not generate the initial excitation. But the reciprocity condition does not require the reproduction of initial conditions; it only requires the equality of forward and backward transmission coefficients for individual modes or ports. It follows from the Lorentz reciprocity theorem that the forward and backward transmission coefficients $T_{12}$ and $T_{21}$ between any two modes of the system are equal. This is identical to the symmetry of the scattering matrix $S=S^T$, a property mentioned but not correctly considered by the authors.

To summarize, the authors have discussed a multimode system without correctly considering the reciprocity properties of transmissions between individual modes of the system. They also erroneously assume time-reversal symmetry to be a necessary condition for reciprocity. Using a non-standard definition of reciprocity, they wrongly claim to have found optical isolation in a purely linear and passive silicon photonic structure.